\documentclass[aps,letterpaper,twocolumn,amsmath,showpacs]{revtex4} 
\usepackage[dvips]{graphicx}
\usepackage{bm}
\begin{document}
\title{Fitting inverse power-law quintessence models using the SNAP satellite}
\date{7 February 2002}
\author{Martin Eriksson}
\email{mate@physto.se}
\author{Rahman Amanullah}
\email{rahman@physto.se}
\affiliation{Department of Physics, Stockholm University, S - 106 91 Stockholm}
\begin{abstract}
We investigate the possibility of using the proposed SNAP satellite in combination with low-$z$ supernova searches to distinguish between $\emph{different}$ inverse power-law quintessence models. If the true model is that of a cosmological constant, we determine the prospects of ruling out the inverse power-law potential. We show that SNAP combined with e.g.~the SNfactory and an independent measurement of the mass energy density to 17\% accuracy can distinguish between an inverse power-law potential and a cosmological constant and put severe constraints on the power-law exponent.
\end{abstract}
\maketitle

%%%%%%%%%%%%%%%%%%%%%%%%%%%%%%%%%%%%%%%%%%%%%%%%%%%%%%%%%%%%%%%%%%%%%%%%%%%%%%
\section{Introduction}
Recent high precision measurements on type Ia supernovae (SNIa), the cosmic microwave background (CMB) and rich galaxy clusters indicate that our Universe is well described by a flat Friedmann-Robertson-Walker (FRW) model with mass energy density $\Omega_m\sim 0.3$ and dark energy density $\Omega_d\sim 0.7$ \cite{sc}. However, our knowledge of the nature of the dark energy is still limited, amounting to the facts that it is smooth on small scales and has negative pressure, therefore causing the Universe to accelerate. The most famous dark energy model which meets these two requirements is vacuum energy due to the cosmological constant $\Lambda$, but equally well-known are the many problems associated with its introduction \cite{ca}. A fairly recent alternative to $\Lambda$ is quintessence, which resembles inflation in that it introduces a new scalar field $Q$ to account for the dark energy. Consequently, the equation-of-state parameter $w_Q = p_Q / \rho_Q$ is in general time-varying, whereas $w_\Lambda = -1$ always. At the present epoch, the field sits still or rolls slowly down its potential which results in a negative equation of state, i.e. negative pressure. A key feature of many quintessence models is that the equations of motion have attractor-like solutions, which ensures that a wide range of initial conditions give the same late-time evolution of the field. Although this behaviour solves one of the fine-tuning problems of the cosmological constant, the tuning of initial conditions, it does not answer the question why $\Omega_m\sim \Omega_d$ today \cite{we}. 

The prospects of using the proposed SuperNova/Acceleration Probe (SNAP) \cite{sn}, a two-meter satellite telescope dedicated to the search and follow-up of supernovae, as a tool to probe the nature of dark energy has been examined in several papers (see \cite{ma,ch,as,hu,go,wa,ge} and references therein). For example, Goliath et al. show that given an independent high-precision measurement of $\Omega_m$, SNAP can constrain the parameters of a linear equation of state $w_Q(z) = w_Q^{(0)} + w_Q^{(1)} z$ to within $\pm 0.04$ and ${}^{+0.15}_{-0.17}$ respectively at the one-sigma level, assuming a flat Universe. Moreover, Weller \& Albrecht conclude that SNAP will be able to distinguish some quintessence models from a pure cosmological constant. In this paper, we investigate the possibility to differentiate between different parameters in one and the same potential. As our data we will use simulated SNIa magnitude-redshift measurements corresponding to one year of SNAP data and 300 events from low-$z$ supernova searches such as the SNfactory \cite{al}.

One of the most common quintessence models in the literature is the simple inverse power-law potential introduced by Ratra and Peebles \cite{ra} and reanalysed by Steinhardt et al. \cite{st}, 
\begin{equation}
	V = \frac{M^{4 + \alpha}}{Q^\alpha}.
\end{equation}
Here $M$ is a mass parameter which is fine-tuned to give the right $\Omega_d$ today when $Q$ is of the order of unity in Planck units \cite{st}. The exponent $\alpha$ is the parameter we want to estimate. It determines the value of $w_Q$ today, with smaller $\alpha$ giving larger negative $w_Q$. In the limit $\alpha \rightarrow 0$ we retrieve the cosmological constant with $\rho_\Lambda = M^4$. Although there is some theoretical motivation for such a potential from supersymmetric QCD (see \cite{nu} and references therein), the main reason for its introduction is phenomenological. In addition to the attractor-like solutions mentioned above, potentials of this type produce an equation-of-state parameter $w_Q$ which automatically decreases to a negative value at the onset of matter domination. Since the energy density of any component evolves as $\rho_i \propto a^{-3(1+w_i)}$ ($a$ being the scale factor), this means that while quintessence starts out as a subdominant contribution, it will eventually come to dominate the Universe.

%%%%%%%%%%%%%%%%%%%%%%%%%%%%%%%%%%%%%%%%%%%%%%%%%%%%%%%%%%%%%%%%%%%%%%%%%%%%%
\section{Method}
\subsection{Basic equations}
We assume a spatially flat FRW Universe with zero cosmological constant and set the Planck mass to unity. For a homogeneous and minimally coupled scalar field, the equations of motion are then given by
\begin{subequations}
\label{eq:motion}
\begin{eqnarray}
	\ddot{Q} + 3H \dot{Q} + V'(Q) = 0, \\
	H^2 = \frac{8 \pi}{3} (\rho_Q + \rho_B), 
\end{eqnarray}
\end{subequations}
where $H$ is the Hubble parameter, $\rho_B$ is the background energy density and dot and prime represent differentiation with respect to time and $Q$ respectively. These equations can be solved numerically for the redshift dependence of the equation of state parameter
\begin{equation}
	w_Q = \frac{\frac{1}{2} \dot{Q}^2 - V}{\frac{1}{2} \dot{Q}^2 + V},
\end{equation} 
independently of the value of the Hubble constant $H_0$. 

The apparent magnitude $m$ of a supernova is $w_Q$-dependent through the luminosity distance $d_L$,
\begin{equation}
	m = \mathcal{M} + 5 \log (H_0 d_L).
\end{equation}
Here $\mathcal{M}$ is a constant which depends on $H_0$ and the absolute magnitude of the supernova which we assume not to evolve with redshift. The second term however, is $H_0$-independent and is given by 
\begin{eqnarray}
	H_0 d_L(z) &=& (1 + z) \times \nonumber \\
	&\times& \int_0^z dz' \left[ \Omega_m (1 + z')^3 + \Omega_Q f(z') \right]^{-\frac{1}{2}}, \\
	f(z) &=& \exp \left( 3 \int_0^{z} dz' \frac{1 + w(z')}{1 + z'} \right).
\end{eqnarray}
At low redshifts, $H_0 d_L \sim z$ for any cosmology which means that low-$z$ events can be used to measure $\mathcal{M}$ without any prior knowledge of the cosmological parameters in general and $H_0$ in particular. This is why $\mathcal{M}$ is sometimes referred to as the Hubble-constant-free absolute magnitude.

\subsection{Simulating supernovae}
We use the SuperNova Observation Calculator (SNOC) \cite{ar}, a program package developed at Stockholm University, to make Monte Carlo simulations of magnitude-redshift datasets attainable with SNAP and other experiments. The fiducial FRW cosmology we use is parametrized by $\mathcal{M} = -3.4$, $\Omega_m = 0.3$, $\Omega_Q = 1 - \Omega_m = 0.7$  and the inverse power-law exponent $\alpha$ an integer number in the range $0 - 10$ (with the value 0 corresponding to a cosmological constant). 

\begin{figure}[htb]
	\includegraphics[width=.4\textwidth]{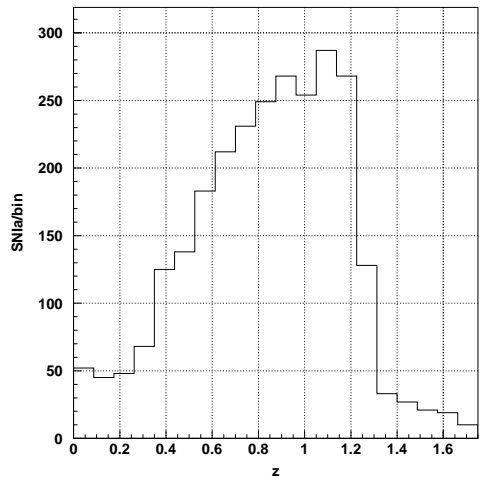}
	\caption{\label{dist} \footnotesize{Simulated SNIa distribution for one year of SNAP data combined with 300 low-$z$ events from e.g.~the SNfactory.}}
\end{figure}

SNAP is expected to observe $\sim 2000$ SNIa per year for three years at redshifts out to $z = 1.7$ \cite{sn}. To this sample we add 300 events at $z < 0.15$ from future low-$z$ supernova searches such as the SNfactory. This is probably a conservative estimate of the amount of low redshift data available at the time of the SNAP launch in 2009. Figure~\ref{dist} shows our simulated SNIa distribution for the combined experiments. The individual measurement error is assumed to be $\sigma = 0.15$ mag, including the intrinsic spread of supernova magnitudes but ignoring systematic effects from e.g.~gravitational lensing or dust.

\subsection{Fitting procedure}
We want to fit our simulated magnitude-redshift measurements for $\alpha$ using the method of maximum likelihood. Since we have assumed a flat Universe, the other unknown parameters are $\Omega_m$ and $\mathcal{M}$. The best-fit values are then found by minimizing the negative log-likelihood function $\mathcal{L}$:
\begin{equation}
	\mathcal{L} = \frac{1}{2} \sum_{i = 1}^n \frac{\left[m(\bm{\theta}, z_i) - m(\bm{\theta}_{\text{sim}}, z_i) \right]^2}{\sigma^2},
\end{equation}
where $\bm{\theta} = (\alpha, \Omega_m, \mathcal{M})$ and subscript sim means simulated values. Note that additive terms which only result in a shift in $\mathcal{L}$ have been dropped in the above expression.

The parameter ranges we consider are $0 \le \alpha \le 10.5$, $0.1 \le \Omega_m \le 0.6$ and $-10 \le \mathcal{M} \le 0$. In order to minimize $\mathcal{L}$ numerically, we need to know $m$ and thus $w_Q$ for any given $\bm{\theta}$. Therefore we first solve the equations of motion (\ref{eq:motion}) for the evolution of $w_Q$ using 165 different $(\alpha, \Omega_m)$ combinations in the above ranges and then use linear interpolation to find $w_Q$ for \emph{any} combination of $\alpha$ and $\Omega_m$. We will display our results using confidence regions in the $(\alpha, \Omega_m)$ plane. 

%%%%%%%%%%%%%%%%%%%%%%%%%%%%%%%%%%%%%%%%%%%%%%%%%%%%%%%%%%%%%%%%%%%%%%%%%%%%%
\section{Confidence regions}
\subsection{No priors}
In this section we assume no prior knowledge of $\mathcal{M}$ nor $\Omega_m$. This means that for each grid point in the $(\alpha, \Omega_m)$ plane, we find the value of $\mathcal{M}$ which minimizes the negative log-likelihood function. Naturally, since we are not trying to estimate $\mathcal{M}$, this value is of no importance to us. 

\begin{figure}[htb]
	\includegraphics[width=.4\textwidth]{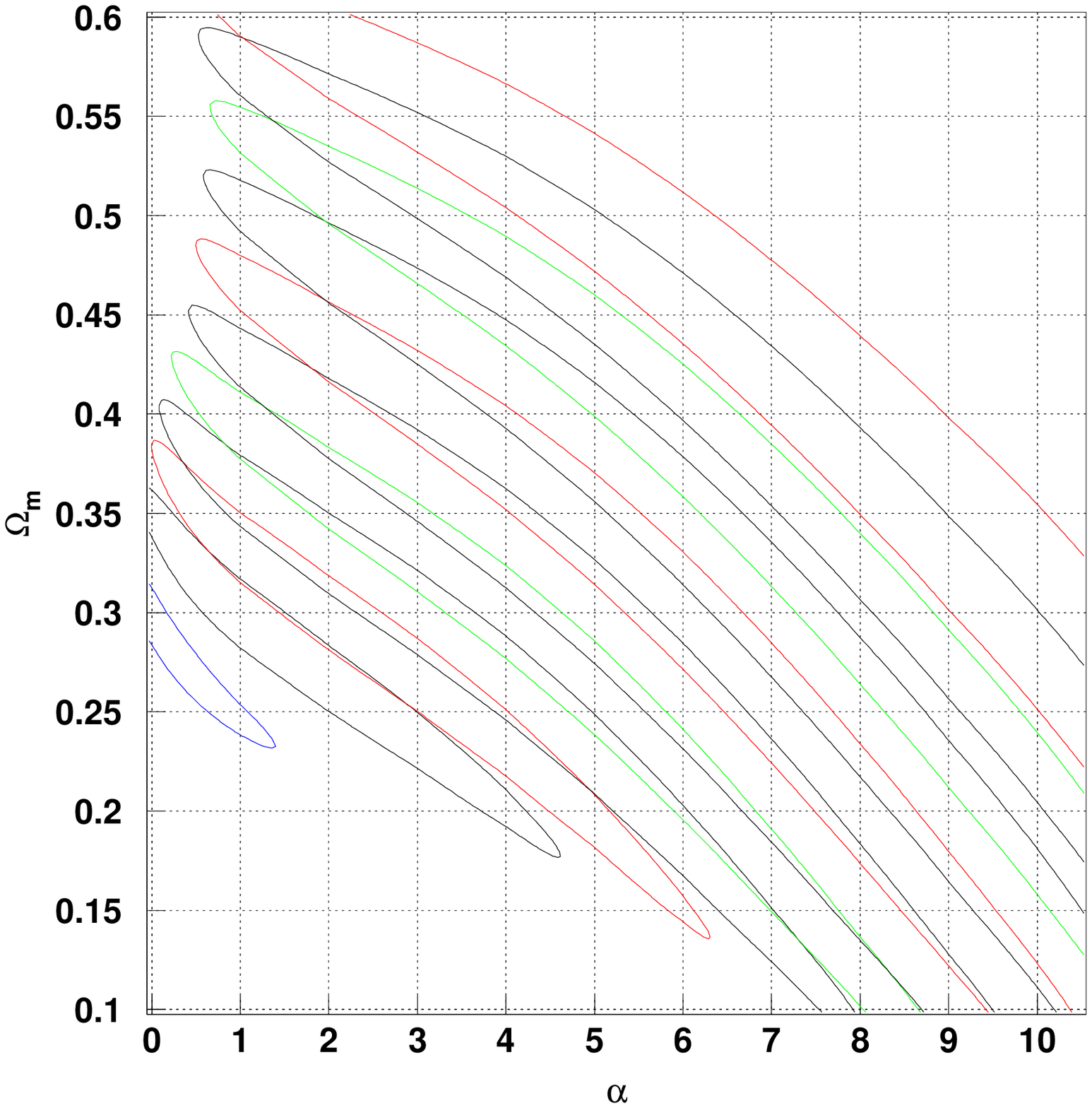}
	\caption{\label{nopriors_snap} \footnotesize{68\% confidence contours for $(\alpha, \Omega_m)$ with one year of SNAP data. The different contours correspond to the simulated fiducial cosmology with $0 \le \alpha \le 10$.}}
\end{figure}

Figure~\ref{nopriors_snap} shows 68\% confidence regions with one year of SNAP data. Different contours correspond to the fiducial cosmology with $0 \le \alpha \le 10$. Due to the strong correlation between $\alpha$ and $\Omega_m$, it is practically impossible to distinguish between different integer $\alpha$ models. The plotted confidence regions are forced to be centered on the true values of the fiducial cosmology, but that will most certainly not be the case for the real confidence region from only one experiment. However, the size and shape of the region will remain approximately the same which means that the region for any $\alpha \ne 0$ may drift to cover almost the entire plane. For $\alpha \le 2$, we can still estimate $\alpha$ to better than ${}^{+2.9}_{-1.4}$ at the one-parameter, one-sigma level. In the case of a cosmological constant and a measured value centered on $\alpha = 0$, we can conclude that $\alpha \le 0.9$ at the one-parameter, one-sigma level.  

\begin{figure}[htb]
	\includegraphics[width=.4\textwidth]{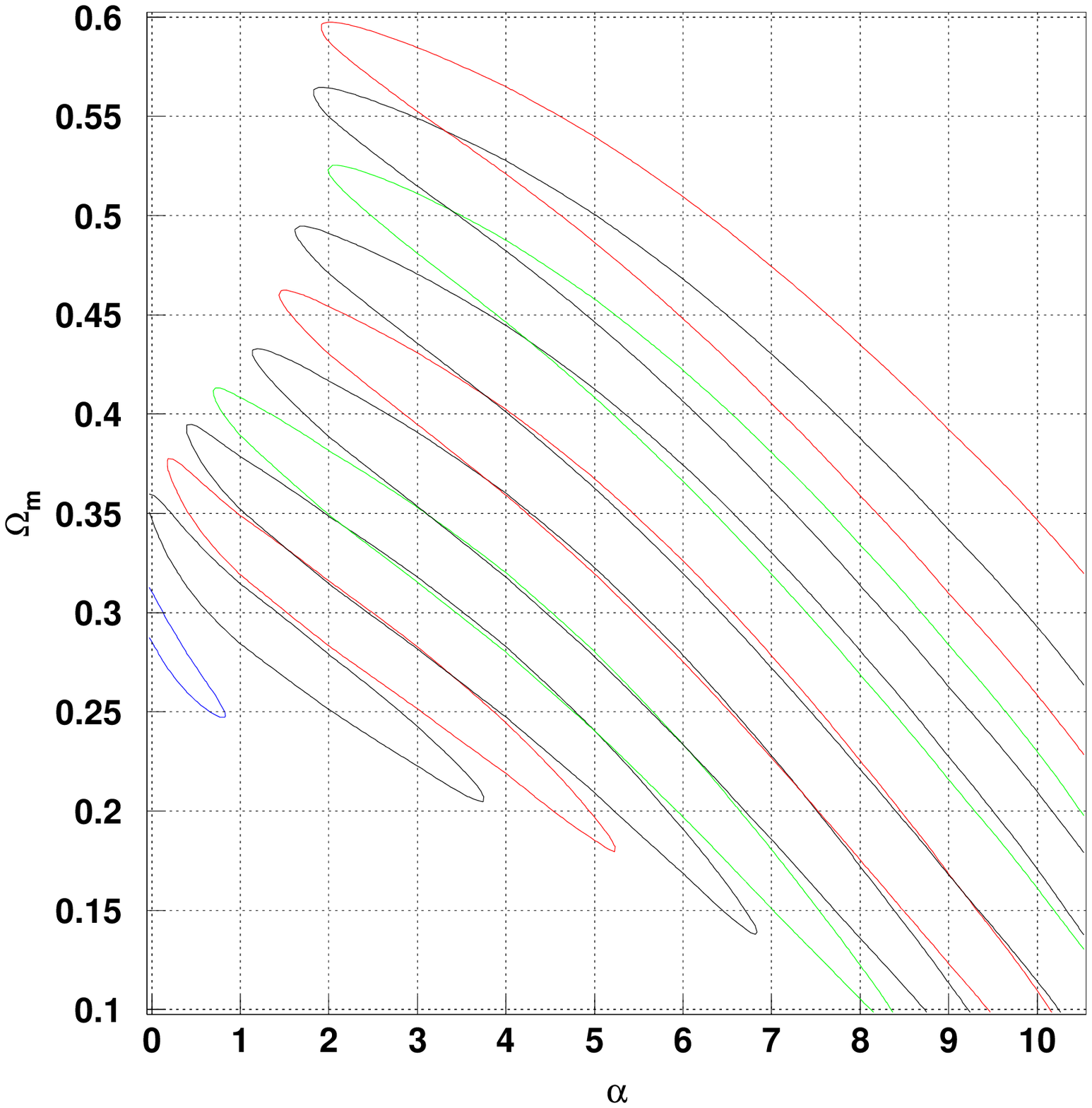}
	\caption{\label{nopriors_lowz} \footnotesize{68\% confidence contours for $(\alpha, \Omega_m)$ with one year of SNAP data and 300 low-$z$ events. The different contours correspond to the simulated fiducial cosmology with $0 \le \alpha \le 10$.}}
\end{figure}

Combining SNAP with e.g.~the SNfactory reduces the error contours somewhat as shown in figure~\ref{nopriors_lowz}. The reason for this was given above; adding low-$z$ supernovae corresponds to prior knowledge of $\mathcal{M}$, something which will become evident later when we assume exact knowledge of $\mathcal{M}$. For $\alpha \le 2$, we can now estimate $\alpha$ to better than ${}^{+2.2}_{-1.2}$ at the one-parameter, one-sigma level. In the case of a cosmological constant and a measured value centered on $\alpha = 0$, we find $\alpha \le 0.6$ at the one-parameter, one-sigma level. However, to break the degeneracy in parameter space we need an independent assesment of $\Omega_m$.

\subsection{Priors on $\Omega_m$ and $\mathcal{M}$}
Our assumption of no prior knowledge of $\Omega_m$ is actually unrealistic, especially since we already assume that the Universe is flat, thereby leaning heavily on the latest results from CMB measurements. These same results, as well as independent observations of the evolution of the number density of galaxy rich clusters and mass estimates of galaxy clusters, all favour a low mass Universe. However, we must be certain that these measurements really are independent of $w_Q$ before using them to invoke prior knowledge of $\Omega_m$. In \cite{ge}, Gerke \& Efstathiou give a brief summary of present and future experiments which may resolve this issue and conclude that at the time of the SNAP launch, $w_Q$-independent measurements of $\Omega_m$ corresponding to a Gaussian prior knowledge with a spread of $\sigma_{\Omega_m} = 0.05$ may actually be a conservative expectation, whereas $\sigma_{\Omega_m} = 0.015$ could be considered optimistic. We note that a recent analysis of 2dF data already gives $\Omega_m = 0.27 \pm 0.06$ \cite{ve}.

\begin{figure}[htb]
	\includegraphics[width=.4\textwidth]{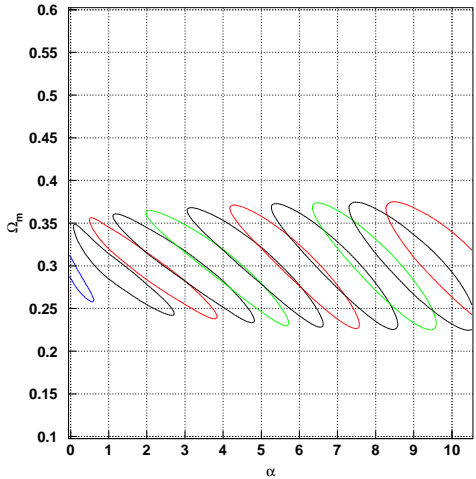}
	\caption{\label{Omm_con_lowz} \footnotesize{68\% confidence contours for $(\alpha, \Omega_m)$ with one year of SNAP data and 300 low-$z$ events, assuming a prior knowledge with $\Omega_m$ Gaussian centered around 0.3 and $\sigma_{\Omega_m} = 0.05$. The different contours correspond to the simulated fiducial cosmology with $0 \le \alpha \le 10$.}}
\end{figure}

\begin{figure}[htb]
	\includegraphics[width=.4\textwidth]{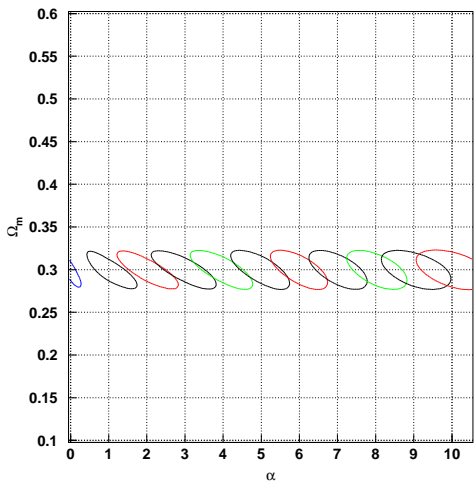}
	\caption{\label{Omm_opt_lowz} \footnotesize{68\% confidence contours for $(\alpha, \Omega_m)$ with one year of SNAP data and 300 low-$z$ events, assuming a prior knowledge with $\Omega_m$ Gaussian centered around 0.3 and $\sigma_{\Omega_m} = 0.015$. The different contours correspond to the simulated fiducial cosmology with $0 \le \alpha \le 10$.}}
\end{figure}

The result of imposing these priors respectively is shown in figures~\ref{Omm_con_lowz} and ~\ref{Omm_opt_lowz}. With $\sigma_{\Omega_m} = 0.05$, we can determine $\alpha$ to better than ${}^{+1.3}_{-1.3}$ at the one-parameter, one-sigma level for all $\alpha$. In the case of a cosmological constant and a measured value centered on $\alpha = 0$, we can conclude that $\alpha \le 1.2$ at the one-parameter, three-sigma level. If we instead use the optimistic spread of $\sigma_{\Omega_m} = 0.015$, we can determine $\alpha$ to better than ${}^{+0.6}_{-0.7}$ at the one-parameter, one-sigma level for all $\alpha$ and in the case of a cosmological constant and a measured value centered on $\alpha = 0$, we can conclude that $\alpha \le 0.6$ at the one-parameter, three-sigma level. 

Finally, we have performed the same analysis assuming exact knowledge of $\mathcal{M}$. This scenario is admittedly utopian but it serves to illustrate the full potential of the SNAP satellite when combined with other experiments. Moreover, there are plans for future experiments which could provide values of $\mathcal{M}$ with a extraordinary precision. The ESA satellite Gaia for example, planned to be launched before 2012, is expected to find roughly $10^5$ low-$z$ supernovae of all types within 4 years \cite{ga}. 

\begin{figure}[htb]
	\includegraphics[width=.4\textwidth]{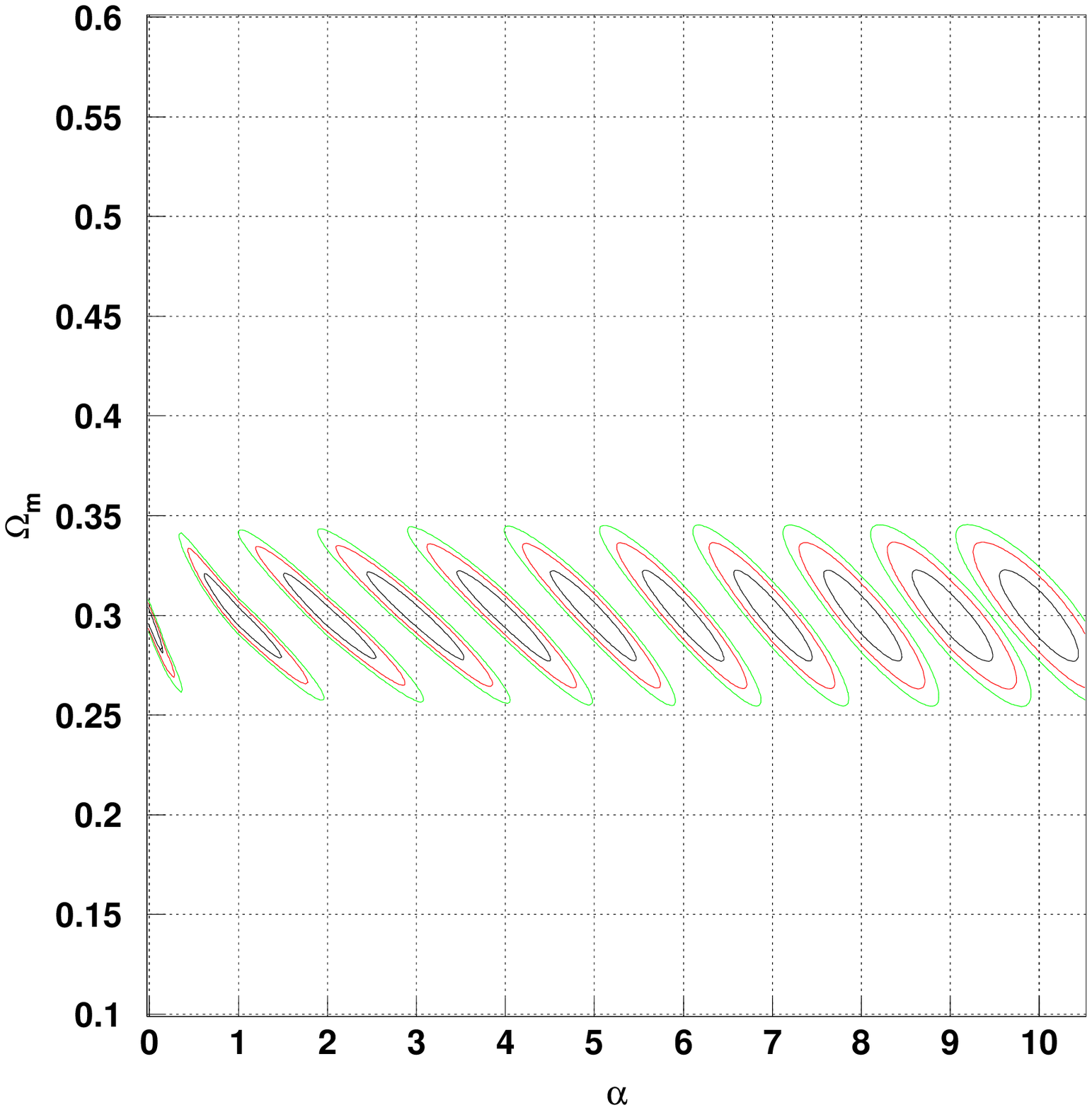}
	\caption{\label{Mfix_Omm_opt_snap} \footnotesize{68\%, 95\% and 99\% confidence contours for $(\alpha, \Omega_m)$ with one year of SNAP data assuming exact knowledge of $\mathcal{M}$ and a prior knowledge with $\Omega_m$ Gaussian centered around 0.3 and $\sigma_{\Omega_m} = 0.015$. The different contours correspond to the simulated fiducial cosmology with $0 \le \alpha \le 10$.}}
\end{figure}

Figure~\ref{Mfix_Omm_opt_snap} shows 68\%, 95\% and 99\% confidence regions for $(\alpha, \Omega_m)$ assuming exact knowledge of $\mathcal{M}$ and $\sigma_{\Omega_m} = 0.015$. It is clear that such a scenario would enable pinpoint precision in the determination of $\alpha$ and in the case of a cosmological constant the ability to rule out $\alpha \ge 1$ at the 99\% confidence level.

%%%%%%%%%%%%%%%%%%%%%%%%%%%%%%%%%%%%%%%%%%%%%%%%%%%%%%%%%%%%%%%%%%%%%%%%%%%%%
\section{Conclusions}
We have investigated the possibility of using future SNIa observations to constrain the exponent $\alpha$ of the inverse power-law quintessence potential. We do not use any parametrization of the equation of state nor any of the ``standard practices'' which according to \cite{ma2} may result in substantial estimation errors. In spite of the outstanding redshift range and measurement precision of the SNAP satellite, as well as the relatively accurate determination of $\mathcal{M}$ possible with e.g.~the SNfactory, it is clear that an independent measurement of the mass energy density $\Omega_m$ is needed in order to make any firm predictions of the value of $\alpha$. 

However, armed with a Gaussian prior on $\Omega_m$ with a spread of $\sigma_{\Omega_m} = 0.05$, we can determine $\alpha$ to better than ${}^{+1.3}_{-1.3}$ at the one-parameter, one-sigma level for $1 \le \alpha \le 10$. In the case of a cosmological constant and a measured value centered on $\alpha = 0$, we can conclude that $\alpha \le 1.2$ at the one-parameter, one-sigma level. In order to take full advantage of the SNAP satellite's capabilities, we would need an even more precise measurement of $\mathcal{M}$ than possible with 300 low-$z$ SNIa. With such a measurement, the determination of $\alpha$ would be very accurate indeed. 

One can argue that the inverse power-law potential is not the most favoured using current data because it does not provide low enough values of $w_Q$. Sticking to integer $\alpha$ one finds $w_Q \gtrsim -0.8$ today, and values as high as $\alpha = 10$ are already ruled out with present data. We have included them here to illustrate the method used and the possible accuracy of the SNAP satellite. In view of this, the most important aspect of this analysis is the ability to rule out the inverse power-law potential. As such we have shown that the method is effective and it is immediately generalizable to other potentials which mimic a cosmological constant for some choice of parameter(s). Our fitting procedure may also be used to rule out the inverse power-law potential in the case of other quintessence potentials. This is work in progress. 

%%%%%%%%%%%%%%%%%%%%%%%%%%%%%%%%%%%%%%%%%%%%%%%%%%%%%%%%%%%%%%%%%%%%%%%%%%%%%
\begin{acknowledgments}
We are greatful to Lars Bergstr\"om, Ariel Goobar, Edvard M\"ortsell, Christian Walck and Nelson Nunes for helpful discussions.
\end{acknowledgments} 

%%%%%%%%%%%%%%%%%%%%%%%%%%%%%%%%%%%%%%%%%%%%%%%%%%%%%%%%%%%%%%%%%%%%%%%%%%%%%


\begin{thebibliography}{99}
\bibitem{sc} N. A. Bahcall, J. P. Ostriker, S. Perlmutter and P. J. Steinhardt, \emph{Science} \textbf{284}, 1481 (1999).
\bibitem{ca} S. M. Carroll, W. H. Press and E. L. Turner, \emph{Annu. Rev. Astron. Astrophys.} \textbf{30}, 499-542 (1992).  
\bibitem{we} S. Weinberg, \eprint{astro-ph/0005265}.
\bibitem{sn} SuperNova/Acceleration Probe (SNAP) 2000, \url{http://snap.lbl.gov}
\bibitem{ma} I. Maor, R. Brustein and P. J. Steinhardt, \emph{Phys. Rev. Lett.} \textbf{86} 6 (2001).
\bibitem{ch} T. Chiba and T. Nakamura, Phys. Rev. D \textbf{62}, 121301.
\bibitem{as} P. Astier, \emph{Phys. Lett. B} \textbf{500}, 8 (2001).
\bibitem{hu} D. Huterer and M. S. Turner, Phys. Rev. D \textbf{64}, 123527 (2001).
\bibitem{go} M. Goliath, R. Amanullah, P. Astier, A. Goobar and R. Pain, \eprint{astro-ph/0104009}.
\bibitem{wa} J. Weller and A. Albrecht, \eprint{astro-ph/0106079}.
\bibitem{ge} B. F. Gerke and G. Efstathiou, \eprint{astro-ph/0201336}.
\bibitem{al} Aldering et al., Supernova Factory Webpage (\url{http://snfactory.lbl.gov}).
\bibitem{ma2} I. Maor, R. Brustein, J. McMahon and P. J. Steinhardt, astro-ph/0112526.
\bibitem{ra} B. Ratra and P. J. E. Peebles, \emph{Phys. Rev. D} \textbf{37}, 3406 (1988).
\bibitem{st} R. R. Caldwell, R. Dave and P. J. Steinhardt, \emph{Phys. Rev. Lett.} \textbf{80}, 1582 (1998); I. Zlatev, L. Wang and P. J. Steinhardt, \emph{Phys. Rev. Lett.} \textbf{82}, 896 (1999).
\bibitem{nu} E. J. Copeland, N. J. Nunes and F. Rosati, Phys. Rev. D \textbf{62} 123503 (2000).
\bibitem{ar} A. Goobar et al., in preperation (2002).
\bibitem{ve} L. Verde et al., \eprint{astro-ph/0112161}.
\bibitem{ga} GAIA homepage, \url{http://sci.esa.int}
\end{thebibliography}
\end{document}